\documentstyle[twoside]{memsait}
\input epsf.sty

\begin{opening}
\title{A new unifying model for X--ray and radio--selected BL Lacs}

\author{G. Fossati$^1$, A. Celotti$^1$, G. Ghisellini$^2$, L. Maraschi$^2$}
\institute{$^1$International School for Advanced Studies, Trieste, Italy\\
$^2$Osservatorio Astronomico di Brera--Merate, Milano, Italy}
\date{} 
\end{opening}

\begin{document}

\oddpagefooter{}{}{} 
\evenpagefooter{}{}{} 
\ 
\bigskip

\begin{abstract}

We discuss alternative interpretations of the differences in the Spectral
Energy Distributions (SEDs) of BL Lacs found in complete Radio or X--ray
surveys.

In order to explain the different properties of radio and X-ray selected
within the assumption of a single population we propose a new approach 
(Fossati et al. 1997). 
The model is based on the idea that the BL Lacs constitute a ``parameterized
family'' and that the physical parameter which governs the shape of the SEDs, 
is (or is associated with) the bolometric luminosity. 
Assuming an empirical relation between spectral shape and luminosity the 
observational properties of the two surveys, including the redshift 
distributions, can be reasonably well reproduced.

\end{abstract}

\section{Different energy cut--off hypothesis}

BL Lacs have been almost exclusively discovered through radio or X--ray
surveys. However the properties of objects selected in the two spectral
bands are systematically different, posing a question as to whether there
are two ``types'' of BL Lacs.  
The first difference to be recognized and
perhaps still the most striking is the shape of the SED. The differences
show up in broad band spectral indices and color--color diagrams, e.g.
$\alpha_{RO}$ vs $\alpha_{OX}$ (Sambruna, Maraschi and Urry 1996). 

On the basis of the X-ray to radio flux ratio we adopt an objective criterion 
separating two (putative) classes of objects: we define XBLs the objects
with $\log (F_{\rm 1keV}/F_{\rm 5GHz}) \ge -5.5$ and RBLs those with a
ratio smaller than this dividing value, where the fluxes are monochromatic and
expressed in the same units. 

The spread in spectral shapes was originally attributed to orientation
effects, associated with different strength of the relativistic boosting 
and widths of the beaming cones of the
radio and X--ray radiation emitted by a relativistic jet, in the sense
of weaker boosting and wider cone for X--ray than for radio emission
(Maraschi et al. 1986). 
The different beaming affecting the various bands could be due to
an accelerating (Ghisellini \& Maraschi 1989) or an increasingly
collimated jet (Celotti et al. 1993).

Substantial progress in the knowledge of the broad band spectra of BL
Lacs suggests that the 
SEDs of XBLs differ from those of RBLs mainly in having a (synchrotron) 
spectral cut--off (or break) at much higher frequency. 
Sambruna et al. (1996) showed that it is difficult to model the detailed 
transition from an XBL to an RBL like in terms of orientation only and 
suggested rather a continuous change in the physical parameters of the jet.

Giommi \& Padovani (1994) introduced then the idea that BL Lacs are a single
population of objects whose SED can be characterized phenomenologically by
the distribution of the values of the frequency at which the peak in the
emission occurs (i.e. the peak in the $\nu
F_{\nu}$ representation of the broad band energy distribution) for the
putative synchrotron component.
In particular they propose that a single luminosity function in the 
radio band describes the full BL Lac population. 
For each radio luminosity X--ray bright BL Lacs (i.e. XBLs) 
are intrinsically a minority following a fixed (luminosity independent)
distribution of X--ray to radio flux ratios.
According to this approach the intrinsic fraction of the two types of BL
Lac would be objectively reflected in radio surveys.

In addition to this ``radio leading'' model we considered the ``symmetric'' 
alternative, ``X--ray leading'', namely that the X-ray luminosity function 
represents the whole BL Lac population.  
In this case the objective survey would be in X--ray band. 
Starting from these hypothesis, one can again derive the statistical properties
of a sample of BL Lacs simply by assuming a radio (X--ray) luminosity 
function and a probability distribution, $\cal P$, of the X--ray to radio 
luminosity ratio. 

Given the above assumptions, we computed the predictions of both models. We
used a Monte Carlo technique to simulate the distribution of sources in
space and luminosity. This method is very convenient because it allows us
to store ``single source'' attributes and not only to compute sample
integrated average properties. 

The predictions were compared in detail with the best BL Lac samples
nowadays available, i.e. the 1 Jy sample (Stickel et al. 1991) and the
one derived from the {\sl Einstein} Slew survey sample (Perlman et al. 1996).
The quantities compared with the predictions of the models are the number 
of sources of the two types and their respective average radio and X--ray 
luminosities, and also the redshift distribution, for both surveys.

In terms of source numbers and average luminosities both models are in 
fairly good agreement with the observed properties of the reference samples, 
showing only minor discrepancies which however correspond to quantities
only weakly constrained due to poor statistics of the ``real'' samples. 
The redshift distribution instead appears badly reproduced by both models. 
In particular they are not able to account for the flatness of the ``$z$''
distribution of RBLs in the 1Jy sample (Fossati et al. 1997).

\section{Unified bolometric approach} 

Because of the implications of these issues for physical models of
relativistic jets and the understanding of the physical conditions within
the emission region, we tried a different approach.

Partly based on the observational indication of a 
possible link, along a continuous sequence, between the
SED shape and the source luminosity (e.g. Sambruna et al. 1996).
we propose that XBLs and RBLs are
different representatives of a {\it spectral sequence} that can 
be described in terms of a single parameter. 
We identify this fundamental quantity with the 
{\it bolometric luminosity of the synchrotron component}, $L_{\rm bol,sync}$.
Both ``flavours'' of BL Lac objects share the same bolometric luminosity
function and the SED properties {\it strongly} depend on it.

The main positive feature of this approach is that it offers a more direct
interpretation in terms of the physical properties of BL Lacs.
In fact:
(a) the assumptions are largely independent of the biases of the observed 
statistical samples;  
(b) there is a more direct connection between the parameters of solutions 
``acceptable'' from a statistical point of view and the physical conditions 
of the emitting plasma.

The relation between the bolometric luminosity and the SED has
been based on observed trends.  More luminous objects seem to have
RBL--type spectral properties, with the peak of the energy distribution in
the mm--IR range and Compton dominated soft X--ray spectra. 
Therefore, if we characterize the SED with the
frequency at which the (synchrotron) energy distribution has a maximum, a
fundamental (inverse) relation must exist between the bolometric
luminosity and this frequency. 

Interestingly, note that the different redshift distributions of the two
kinds of BL Lacs could be a natural outcome of this scenario.  XBLs
objects come from the lower (and richer) part of the luminosity function
and so they would dominate at low redshift, but they would disappear at
large distances despite the increase of the available volume. On the
contrary, RBLs, even though coming from the poorer part of the luminosity
function, would become predominant at higher redshifts, being still
detectable. 

In order to reproduce the basic features of the observed SEDs, we
consider a simple ``geometric'' parameterization. 
The synchrotron radio to soft X--ray component is represented with a power 
law in the radio domain, smoothly connecting with a parabolic branch ranging 
up to $\nu_{\rm peak}$. Beyond $\nu_{\rm peak}$ the synchrotron component 
steepens parabolically.
The hard X--ray Compton component is simply represented with 
a single power law.
This schematic description of the SEDs well reproduces the basic 
properties of the observed BL Lacs broad band spectra.

We then specify the relation between $\nu_{\rm peak}$ 
and the value of $L_{\rm bol,sync}$. 
This is the {\it key} physical relation of the proposed model. 
For simplicity we consider a simple power law dependence:  
$\nu_{\rm peak} \propto L_{\rm bol,sync}^{-\eta}$ ($\eta > 0$).

Despite the number of parameters, there is only a limited freedom in the
choice of their values, which are tightly constrained by observations. 

We adopt a luminosity function for $L_{\rm bol,sync}$ inspired to that
calculated by Urry \& Padovani (1995) (see their Fig.~13) for the 1 Jy and
EMSS BL Lacs. 

\medskip\noindent
Even a ``first attempt'' set of input model parameters gives a
surprisingly good results.
Basically, all observational quantities are correctly predicted by the 
model, except an excessive radio luminosity for RBL detected in radio surveys.
Furthermore, a very positive consequence of this scenario is that it
implies a qualitatively correct redshift distribution.
There is still a problem with RBLs in the radio survey. 
In fact, even though the shape of the distribution is correctly flat, 
and this is the major difficulty of the two other models, it extends 
beyond $z=1$. 
This is likely connected with the excessive average radio luminosity of RBLs. 
However, we suggest that the excess of objects predicted at higher redshift 
(and/or their overestimated radio power) could be understood in an even 
broader unifying picture. 

\section{Discussion and conclusions} 

We propose (Fossati et al. 1997) a new unified bolometric model, whose key 
feature  is the link of the bolometric luminosity with the energy of the
synchrotron cut--off. 
This scenario is based on a schematic parameterization of the BL Lac SED,
constrained by observational trends, and adopts a semi--empirical bolometric 
luminosity function.
Despite the rigid formulation of the one--to--one correspondence between
the SED properties and the luminosity, in this scenario all the main
observational data can be successfully reproduced. 
The only discrepancy of the model seems to be the prediction of a radio
luminosity for RBLs detected in radio surveys higher than observed, and
the related problem with their redshift distribution. 

We note, however, that Highly Polarized Quasars (HPQ) show an interesting 
continuity of properties with those of BL Lacs (e.g. Sambruna et al. 1996; 
Comastri et al. 1997). 
There seems to be a remarkable progression in properties, in a luminosity 
sequence XBL$\to$RBL$\to$HPQ, which is also a sequence of increasing 
importance of emission lines. 
In this picture, if a fraction of the RBL with the highest bolometric 
luminosities were indeed HPQ, this fact could explain the 
discrepancies of the model in the predictions of the high radio luminosity 
and redshift distribution. 

As already stressed, a very interesting aspect of the ``bolometric''
approach is that it is based on the relative dependence of two quantities
which are directly related to the properties of the emitting
plasma, namely the luminosity and the cut--off of the synchrotron
spectrum.  
One can therefore speculate on the physical origin of this dependence. 
Probably the most simple interpretation of this link is that it is
related to the particle cooling. The more radiation is emitted
the more particles loose energy, with consequent decrease of the cut--off
frequency (which depends the maximum energy of the emitting particles).

\end{document}